%% file: entropy_contours.tex
\documentclass[%
 reprint,
superscriptaddress,
nofootinbib,
 amsmath,amssymb,
 aps,
prl,
twocolumn,
]{revtex4-2}

\usepackage{graphicx}
\usepackage{dcolumn}
\usepackage{bm}
\usepackage{bigints}

\newcommand{\pbp}{\left\langle \bar{\psi} \psi \right\rangle}
\newcommand{\MeV}{\, \rm{MeV}}

\begin{document}

\title{Lattice QCD constraints on the critical point from an improved precision equation of state}

\author{Szabolcs Bors\'anyi}
\affiliation{University of Wuppertal, Department of Physics, Wuppertal D-42119, Germany}

\author{Zolt\'an Fodor}
\affiliation{Pennsylvania State University, Department of Physics, State College, PA 16801, USA}
\affiliation{Pennsylvania State University, Institute for Computational and Data Sciences, State College, PA 16801, USA}
\affiliation{University of Wuppertal, Department of Physics, Wuppertal D-42119, Germany}
\affiliation{Institute  for Theoretical Physics, ELTE E\"otv\"os Lor\' and University, P\'azm\'any P. s\'et\'any 1/A, H-1117 Budapest, Hungary}
\affiliation{J\"ulich Supercomputing Centre, Forschungszentrum J\"ulich, D-52425 J\"ulich, Germany}

\author{Jana N. Guenther}
\affiliation{University of Wuppertal, Department of Physics, Wuppertal D-42119, Germany}

\author{Paolo Parotto}
\affiliation{Dipartimento di Fisica, Universit\`a di Torino and INFN Torino, Via P. Giuria 1, I-10125 Torino, Italy}

\author{Attila P\'asztor}
\affiliation{Institute  for Theoretical Physics, ELTE E\"otv\"os Lor\' and University, P\'azm\'any P. s\'et\'any 1/A, H-1117 Budapest, Hungary}

\author{Claudia Ratti}
\affiliation{
 Department of Physics, University of Houston, Houston, TX 77204, USA
}

\author{Volodymyr Vovchenko}
\affiliation{
 Department of Physics, University of Houston, Houston, TX 77204, USA
}

\author{Chik Him Wong}
\affiliation{University of Wuppertal, Department of Physics, Wuppertal D-42119, Germany}

\date{\today}

\begin{abstract}
In this Letter we employ lattice simulations to search for the critical point of 
quantum chromodynamics (QCD). We search for the onset of a first order QCD transition 
on the phase diagram by following contours of constant entropy density from imaginary to 
real chemical potentials under conditions of strangeness neutrality.
We scan the phase diagram and investigate whether these contours meet 
to determine the probability that the critical point is located in a certain region 
on the $T-\mu_B$ plane.
To achieve this we introduce a new, continuum extrapolated equation of state at zero density
with improved precision using lattices with $N_\tau=8,10,12,16$ timeslices,
and supplement it with new data at imaginary chemical potential.
The current precision allows us to exclude, at the 
$2\sigma$ level,  the existence of a critical point at $\mu_B < 450$~MeV.
\end{abstract}

\maketitle

\section{\label{sec:level1}Introduction}

The study of the phase diagram of quantum chromodynamics (QCD) is an extremely 
active field of research, which encompasses conditions ranging from high energy 
heavy ion collisions to the mergers of neutron stars. The major tool for 
\textit{ab initio} investigations of QCD thermodynamics are lattice simulations, 
which have shown that, at vanishing density, the transition separating the hadronic 
phase from the quark gluon plasma (QGP) is a broad crossover~\cite{Aoki:2006we} 
located at around $T=155-160 \MeV$~\cite{Bazavov:2018mes, Borsanyi:2020fev}.

At finite chemical potential $\mu_B$, based on results from several effective 
approaches to QCD, it is conjectured that the crossover transition turns into a 
first order line at a critical endpoint~(see e.g. \cite{Bzdak:2019pkr,Du:2024wjm}). 
Lattice QCD at finite density suffers from a complex action problem, which 
renders simulations extremely costly. Several indirect approaches have been 
devised to circumvent this issue, most notably Taylor expansion around $\mu_B=0$~\cite{Allton:2003vx,Allton:2005gk,Kaczmarek:2011zz,Endrodi:2011gv,Borsanyi:2012cr,Bazavov:2017dus,Bonati:2018nut,Bazavov:2018mes,HotQCD:2018pds,Bollweg:2022rps}, 
analytic continuation from imaginary chemical potential~\cite{Bonati:2018nut,deForcrand:2002hgr,DElia:2002tig,DElia:2007bkz,Cea:2009ba,Bonati:2015bha,Cea:2015cya,Bellwied:2015rza,Vovchenko:2017xad,Vovchenko:2017gkg,Borsanyi:2020fev,Borsanyi:2021sxv,Borsanyi:2022qlh} and reweighting 
techniques~\cite{Barbour:1997ej,Fodor:2001au,Fodor:2001pe,Fodor:2004nz,deForcrand:2002pa,Alexandru:2005ix,Fodor:2007vv,Endrodi:2018zda,Giordano:2020uvk, Giordano:2020roi, Borsanyi:2021hbk,Borsanyi:2022soo}. 
Results from these methods have been obtained for different quantities, like the 
equation of state, fluctuations of conserved charges or the QCD transition line, up to 
chemical potentials as large as $\mu_B \simeq 3.5 \, T$, though no sign of 
criticality has been observed. Notable exceptions are attempts to extract the 
position of the corresponding Lee-Yang edge singularity from lattice data~\cite{Bollweg:2022rps,Basar:2023nkp,Clarke:2024ugt,Adam:2025pii}.

The quest for the QCD critical point was one of the main motivations behind the
extensive Beam Energy Scan program at the Relativistic Heavy Ion Collider, from
which final results will come in the near future.  Although direct experimental
evidence for the existence of a critical point is still missing, both the
BES-I~\cite{STAR:2020tga,STAR:2021iop} and preliminary BES-II
results~\cite{Pandav:CPOD2024} on net-proton cumulants indicate deviations from
non-critical baselines~\cite{Braun-Munzinger:2020jbk,Vovchenko:2021kxx} at the
lowest energies, which might not be explainable without critical effects.  At
the moment, the uncertainties are still sizable, but more precise data and
analysis in the future might clarify the issue.

Several recent predictions for the location of the critical point have been
obtained via functional
methods~\cite{Isserstedt:2019pgx,Fu:2019hdw,Gao:2020fbl,Gao:2020qsj,Gunkel:2021oya,Fu:2021oaw},
holography~\cite{Critelli:2017oub,Hippert:2023bel,Zhu:2025gxo} and the analysis
of experimental
data~\cite{Fraga:2011hi,Lacey:2015yxg,Sorensen:2024mry,Steinheimer:2025hsr}.
Many (but not all) of these predictions based on different approaches 
cluster in a rather small region on the QCD 
phase diagram. However, the systematic
uncertainty of these approaches is hard to estimate.
Hence, a method with controlled systematics is highly desirable. 

\begin{figure*}[t]
    \centering
    \includegraphics[width=0.475\linewidth]{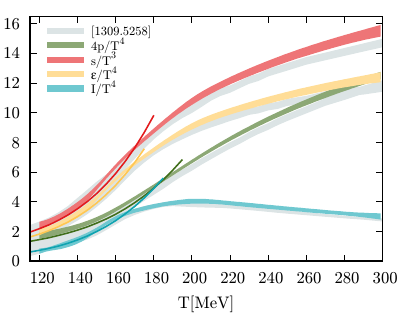}
    \hfill
    \includegraphics[width=0.515\linewidth]{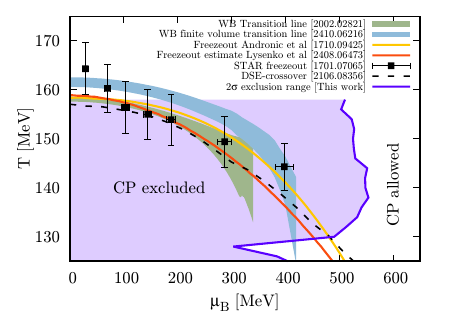}
    \caption{
Left: pressure (green), entropy (red), energy density 
(yellow) and trace anomaly (cyan) as functions of the temperature. The gray bands 
show our previous results from Ref.~\cite{Borsanyi:2013bia}, while the curves at low temperature are the HRG model results. 
The deviations between the two sets of data at high temperature are due to the inclusion of the charm quark in our new results, as opposed to the 2+1 flavor result from Ref.~\cite{Borsanyi:2013bia}.
Right: The QCD phase diagram in the $T-\mu_B$ plane with the $2\sigma$ exclusion range obtained in this work.
We also show the chiral transition line from the continuum extrapolated lattice simulations 
in a large volume \cite{Borsanyi:2020fev}, as well as a recent result in a smaller system
without continuum extrapolation \cite{Borsanyi:2024xrx} and a prediction based on Dyson-Schwinger equations \cite{Gunkel:2021oya}.
Besides the STAR result for chemical freeze-out parameters for collision energies $\sqrt{s} = 200 - 7.7$~GeV \cite{STAR:2017sal} 
and its parametrization \cite{Andronic:2017pug}, we add
a recent lower bound estimate based on a fixed energy/particle number ratio within the HRG framework \cite{Lysenko:2024hqp}.
    \label{fig:exclProb}
}
\end{figure*}

In Ref.~\cite{Shah:2024img} a novel method, based solely on lattice results, was 
proposed to estimate the location of the critical point by studying contours of 
constant entropy density on the QCD phase diagram. The idea is to employ the 
entropy contours to search for the onset of a first order regime: if different 
contours meet somewhere in the phase diagram, it means that the entropy has become 
multi-valued, hence the transition has turned first order. By determining where the 
first order regime is reached, the location of the critical point can be estimated.
In Ref.~\cite{Shah:2024img}, the entropy contours were constructed based on
existing lattice results at vanishing chemical potential for the 
entropy density~\cite{Borsanyi:2013bia} and second order baryon 
susceptibility~\cite{Borsanyi:2021sxv}.

In this work, we generalize this method in more than one way.
To reach the necessary precision to bear predictive power, we perform a new
computation of the equation of state of QCD at zero density, with significantly
reduced uncertainties compared to existing results.
Instead of extracting the constant entropy contours from $\mu_B=0$ data,
we extrapolate them from imaginary $\mu_B$.
In addition, we increase the statistics of data sets at the three highest
imaginary chemical potentials on the finest lattice in comparison to our earlier
work \cite{Borsanyi:2022qlh}. While the extrapolation in
Ref.~\cite{Shah:2024img} was done at order $\mathcal{O}(\mu_B^2)$, here we allow for leading
corrections to the linear $\mu_B^2$ dependence. We use the actual statistical correlations
between variables and also consider additional sources of
systematic errors in the analysis.
Besides, the results presented in Ref.~\cite{Shah:2024img} were obtained
for zero strangeness chemical potential, while here we enforce strangeness neutrality.

There are two main results in this Letter. We present an update of the
zero-density equation of state (see tabulated data in the Supplemental
Material).  We also determine an exclusion probability for the existence of a
critical point as a function of the chemical potential. We show both results in
Fig.~\ref{fig:exclProb}.  We obtain our final result by combining our excluded
region  with the available information on the chemical freeze-out line
\cite{Andronic:2017pug,STAR:2017sal,Lysenko:2024hqp}.  To be conservative, we
choose the curve in Ref.~\cite{Lysenko:2024hqp}, which was estimated as a lower
bound for the freeze-out temperature, and thus, also for $T_c$.  We find that,
at the $2\sigma$ level, we can exclude the existence of a critical point below
$\mu_B^{2\sigma}=450$~MeV.

In the following, we discuss the ingredients that have led us to this finding.
We first introduce the update to the zero-density equation of state
\cite{Borsanyi:2013bia} at $\mu_B=0$ in the temperature range $T=120-300$~MeV.
We extend the entropy function to imaginary $\mu_B$ by continuum extrapolating
the imaginary baryon density and its temperature derivative at these $\mu_B$,
maintaining strangeness neutrality. Finally, the contours of constant entropy
are extrapolated to $\mu_B>0$, and we calculate the probability for the entropy
to be multi-valued at each point on the phase diagram,

\section{\label{sec:eosmu0}Equation of state at $\mu_B=0$\protect\\ }

Throughout the analysis, we employ our 4stout staggered action with $N_f=2+1+1$
quarks at physical quark masses. Thus, we compute the QCD pressure including
the dynamical contribution of the charm quark. In Ref.~\cite{Borsanyi:2016ksw}
we already computed the equation of state with this lattice action at high
temperatures, now we complement this to the transition region.  Here the charm
quark plays an insignificant role except for high temperatures, $T \gtrsim
250$~MeV.  Methodically, we proceed as in Ref.~\cite{Borsanyi:2013bia}, where
the 2+1 flavor equation of state was computed earlier by us. 

\begin{figure}
\includegraphics[width=\linewidth]{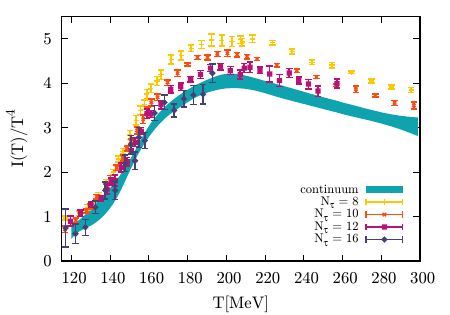}
\caption{Trace anomaly in the continuum (cyan band), as well as at finite 
$N_\tau$ (colored points). 
\label{fig:tracea}}.
\end{figure}

To determine the equation of state at vanishing chemical potential, we employ
the integral method of Ref.~\cite{Engels:1990vr}, whereby the pressure is
obtained through the integration of the trace anomaly:
\begin{equation} \label{eq:press} 
    \frac{p(T)}{T^4} = \frac{p(T_0)}{T_0^4} + \int_{T_0}^{T} \frac{dT^\prime}{T^\prime} \frac{I(T^\prime)}{{T^\prime}^4} \, \, ,
\end{equation}
where $\frac{p(T_0)}{T_0^4}$ is an integration constant. 

The integration constant and the trace anomaly are addressed separately with
designated lattice runs, both are extrapolated to the continuum limit. We do
not use model-dependent input~(such as hadron resonance gas) to fix the
integration constant, or to constrain the continuum limit.

Equation~(\ref{eq:press}) is a multi-dimensional line integral in the space of
the bare parameters of the action -- the gauge coupling $\beta=6/g^2$ and the
fermion masses $m_f$ --  along the line of constant physics $m_f (\beta)$. The
bare parameters of this discretization have been documented in
Ref.~\cite{Bellwied:2015lba}. Throughout this work, we use physical quark
masses on $N_t =8,10,12,16$ lattices.

The trace anomaly can be written as:
\begin{equation}
\frac{I(T)}{T^4} = - T N_\tau^4 \frac{d \beta}{d T} \left( \left\langle - s_G \right\rangle + \sum_f \frac{dm_f}{d\beta} \left\langle \bar{\psi}_f \psi_f  \right\rangle \right) 
\end{equation}
where $d\beta/dT$ is given by the scale setting and the $dm_f/d\beta$ are given
by the line of constant physics.  The expectation values $\left\langle - s_G
\right\rangle$ and $\left\langle \bar{\psi}_f \psi_f  \right\rangle$ are the
gauge action and the chiral condensates, respectively. Both quantities require
additive renormalization, which is carried out by subtracting the vacuum
contribution at the same bare parameters:
\begin{align}\label{eq:rensG1}
\left\langle - s_G \right\rangle &= \left\langle - s_G \right\rangle_T - \left\langle - s_G \right\rangle_0 \, \, , \\\label{eq:renpbp}
\left\langle \bar{\psi}_f \psi_f  \right\rangle &= \left\langle \bar{\psi}_f \psi_f  \right\rangle_T - \left\langle \bar{\psi}_f \psi_f  \right\rangle_0 \, \, .
\end{align}

We performed a large number of zero-temperature simulations to reduce the noise
on the vacuum term in Eq.~\eqref{eq:rensG1}, which is responsible for much of
the error on the trace anomaly itself~\cite{Borsanyi:2010cj}.  We determine the
trace anomaly in the range $T =120 - 300$~MeV by combining a linear continuum
limit in $1/N_\tau^2$ with a spline fit in the temperature. Details on the
statistics and the fit procedure can be found in the Supplemental Material. The
continuum extrapolated trace anomaly is shown in Fig.~\ref{fig:tracea},
together with the results on individual lattices.

The integration constant in Eq.~\eqref{eq:press} is determined in a similar
way, by integrating the chiral condensate in the fermion masses from infinity,
where the pressure vanishes, down to the physical values. In practice, it is
sufficient to start from the mass of the charm quark $m_c$, integrate the
3-flavour theory down to the strange quark mass $m_s$, then integrate the
two-flavour theory down to the light quark mass $m_l$: 
\begin{align} \label{eq:pconst} 
\frac{p(T_0)}{T_0^4} &= \int_{m_s}^{m_l} dm_2 \pbp{\!}_{2} \,  (m_2) \\ \nonumber
& \quad + \int_{\infty}^{m_s} dm_3 \pbp{\!}_{3} \, (m_3) \, \, .
\end{align}
Here, the subscripts $2,3$ refer to the $N_f=2,3$ flavour theories, and the
chiral condensates are the renormalized ones. In this work we choose to
calculate this constant at $T_0 = 185 $~MeV, to minimize the uncertainty on the
equation of state in the transition region.

For the continuum extrapolation of $p(T_0)/T^4_0$ we include $N_\tau=10,12,16$
lattices. Several sources of systematic uncertainties are considered, for a
total of 32 analyses, which are then combined with the histogram method
introduced in Ref.~\cite{Borsanyi:2020mff}. The final result is:
\begin{equation}
    \frac{p(T_0=185 \MeV)}{T_0^4} = 1.371(28) \,\, ,
\end{equation}
where the quoted error includes both statistical and systematic uncertainties.
The precision we reach is approximately a $2\times$ improvement over our
previous result \cite{Borsanyi:2013bia}. Details on the extrapolation and the
systematic analyses can be found in the Supplemental Material.

With Eq.~\eqref{eq:press} we can now determine the pressure, and from it the
other thermodynamic quantities. Trace anomaly, pressure, entropy and energy
density are shown in the left panel of Fig.~\ref{fig:exclProb} as colored
bands, together with our previous results from Ref.~\cite{Borsanyi:2013bia},
shown as gray bands for comparison. The improved precision in both terms of
Eq.~\eqref{eq:press} results in a substantial improvement for the precision of
all quantities, especially near the transition region. Note that our previous
results were obtained with $N_f = 2+1$ quarks: the contribution of the charm
quark becomes non-negligible at around $T \simeq 250 \MeV$.

\begin{figure}\centering
\includegraphics[width=\linewidth]{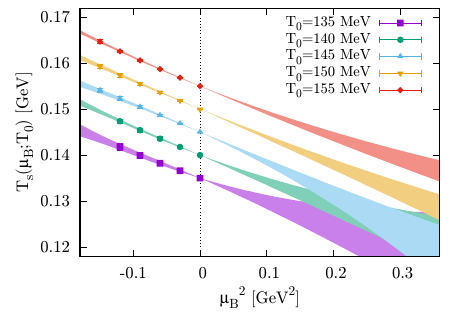}
\caption{Contours of constant (dimensionful) entropy density as fitted rational
functions on continuum extrapolated lattice data.  Each contour is labeled by
its crossing point at $\mu_B=0$: $T_0$. Note that the extrapolation is defined
in the dimensionful $\mu_B^2$ (see also \cite{Shah:2024img}), not $\mu_B/T$.
\label{fig:entr_contours}
}
\end{figure}

\section{Entropy contours}
In this work we construct and follow contours of constant entropy density from
imaginary to real baryon chemical potential.  These contours are defined at
vanishing strangeness density and zero electric charge chemical potential.  At
finite (real or imaginary) $\mu_B$, the entropy density can be written as:
\begin{equation} \label{eq:entr_immu}
    s (T, \mu_B) = s(T, \mu_B=0) + \bigintssss_0^{\mu_B} d\mu_B^\prime \frac{\partial n_B (T, \mu_B^\prime)}{\partial T}  \rm,
\end{equation}
where the first term comes from the continuum extrapolated equation of state
discussed in the previous section, and $n_B(T,\mu_B)$ is the baryon density. We
construct a continuum extrapolation of $n_B(T,\mu_B)$ for imaginary values of
the chemical potential $\mu_B$ by employing simulations at imaginary chemical
potentials on lattices with $N_\tau=10,12$ and $16$ timeslices. We perform a
combined fit in the temperature $T$, the baryochemical potential $\mu_B$ and
the lattice spacing $1/N_\tau^2$. Details on the fit and the systematic
analyses can be found in the Supplemental Material.

From the continuum extrapolated baryon density  we construct the entropy
density at imaginary chemical potential using Eq.~\eqref{eq:entr_immu}. In
order to construct the contours, we proceed as follows. For each temperature
$T_0$, we first calculate the corresponding value $s(T_0)$ of the entropy at
$\mu_B=0$.  For each chemical potential, we then determine the temperature
$T_s(\mu_B,T_0)$ at which the entropy has the desired value $s(T_0)$. The
function $T_s(\mu_B,T_0)$ then yields the contour stemming from $T_0$. By
repeating the procedure for several values of $T_0$, we obtain the contours
shown as points in Fig.~\ref{fig:entr_contours} for $\mu_B^2<0$. We observe
that the dependence of $T_s(T_0, \mu_B)$ on $\mu_B^2$ is remarkably linear to a
high precision. In order to account for deviations from an exact linear
behavior, we consider an extrapolation to real chemical potential based on a
rational ansatz of the form:
\begin{equation}
T_s(\mu_B^2,T_0) = \frac{T_0 + a \mu_B^2}{1 + b \mu_B^2} \, \, .
\end{equation}
The results are shown as bands in Fig.~\ref{fig:entr_contours}.

\begin{figure} 
\centering
\includegraphics[width=\linewidth]{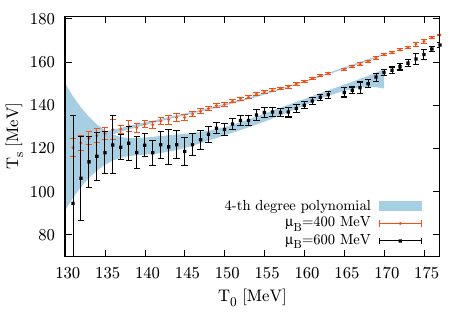}
\caption{\label{fig:TsT0}
The value of $T_s(\mu_B^2,T_0)$ for two fixed values of the baryo-chemical potential. 
These curves develop a negative derivative in the first order regime, and an 
exactly vanishing derivative at the critical point. 
}
\end{figure}

\begin{figure} 
\centering
\includegraphics[width=\linewidth]{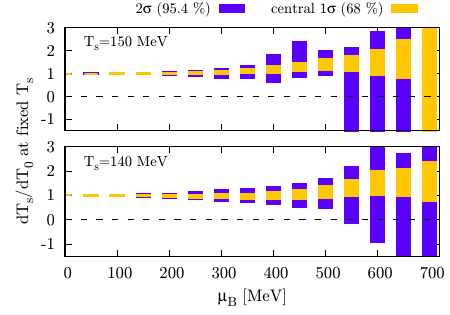}
\includegraphics[width=\linewidth]{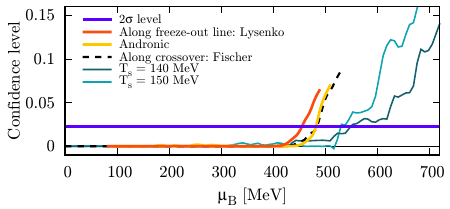}

\caption{
Top: the derivative of the $T_s(T_0)$ curve at fixed chemical potential as a function of $T_0$
(see Fig.~\ref{fig:TsT0}). This quantity is positive for a crossover, negative for 
a first order transition and zero at the critical point. 
Bottom: the confidence level for the $(\partial T_s/\partial T_0)_{\mu_B} < 0$
hypothesis for two fixed values of $T_s$ and along the predicted crossover line
of Ref.~\cite{Gunkel:2021oya}. We also show the confidence levels along the two
freeze-out lines in Fig.~\ref{fig:exclProb}.} \label{fig:conlev}
\end{figure}

In order to determine the onset of a first order transition regime on the phase
diagram, we now study the contours $T_s(\mu_B^2,T_0)$ for different, fixed
values of $\mu_B$. Because the entropy is a monotonic function of the
temperature in the absence of a discontinuity, the curve $T_s(\mu_B^2,T_0)$
will be, at fixed $\mu_B$, monotonic as well. However, if and when the entropy
develops a discontinuity, there will be a regime where the derivative
$(\partial T_s/ \partial T_0)_{\mu_B}$ drops below zero, while being exactly
zero at the critical point. We can see a pattern similar to the expected one in
Fig.~\ref{fig:TsT0}, where $T_s(\mu_B^2,T_0)$ is shown for two values of
$\mu_B$.  At large chemical potentials data are consistent with a zero or
negative derivative within some level of confidence (e.g. $\mu_B=600$~MeV in
Fig.~\ref{fig:TsT0}).

We identify this confidence level for each $T-\mu_B$ pair in the phase diagram.
First, we solve $T=T_s(\mu_B^2,T_0)$ for $T_0$, and then evaluate $(\partial
T_s/ \partial T_0)_{\mu_B}$ at that value.  For a broad range of parameters,
this is at least one or two $\sigma$ larger than zero, indicating that at the
given $T-\mu_B$ position, the entropy is not multi-valued to the given level of
confidence.

We exemplify this in the top panel of Fig.~\ref{fig:conlev}. The two curves
represent two horizontal stripes in the phase diagram, where this analysis was
carried out.  The plot shows the levels that would belong to $1\sigma$ and
$2\sigma$ around the central value, if the distribution was Gaussian. We
translate this to a confidence level in the bottom panel. There we also show
the probability of having a negative slope of $T_s(T_0)$ along two versions of
the chemical freeze-out line \cite{Lysenko:2024hqp,Andronic:2017pug} and a
prediction of the cross-over line \cite{Gunkel:2021oya} in the phase diagram.

The data in Fig.~\ref{fig:conlev} already refer to the full systematic
analysis. In particular, we performed 16 different continuum extrapolations to
arrive at the entropy at imaginary $\mu_B$, and once extrapolated to $\mu_B>0$,
each resulting $T_s$ function was interpolated in 12 different ways.  Though
each of the $12\cdot 16=192$ such analyses provides a Gaussian statistical
error, once their respective cumulative distribution function is averaged, the
result is non-Gaussian.  The main result in Fig.~\ref{fig:exclProb} shows the
chemical potential for a set of fixed temperatures where the confidence level
reaches 0.02275 for $2\sigma$.  At the $2\sigma$ confidence level the lower
bound is  $\mu_B^{2\sigma}=450$~MeV.

\section{Conclusions}

We presented an update of the QCD equation of state at $\mu_B=0$ with improved
precision in comparison to our previous result \cite{Borsanyi:2013bia}.  This
result was obtained with the (2+1+1 flavor) 4stout staggered action up to a
resolution of $N_\tau=16$ to perform a continuum extrapolation.

This allowed us to calculate contours of constant entropy on the QCD phase
diagram in the $T-\mu_B$ plane, by analytic continuation from purely imaginary
chemical potentials using continuum extrapolated lattice QCD simulations in a
large volume: $LT=4$ in temperature units.  Unlike previous estimates based on
lattice simulations, we show results on the phenomenologically relevant
strangeness neutral line.  The contours of  constant entropy in principle allow
one to locate the spinodal region, and thus the onset of a first order
transition at the coveted critical endpoint.  While some of the considered
analyses indeed show a critical endpoint, we have found that after taking into
account all systematic and statistical uncertainties, the location of the
critical endpoint cannot be unambiguously bracketed.  This also means that our
analysis does not provide a clear proof of the existence of such a critical
endpoint. However - for the first time in the literature - we are able to give
reliable lower bounds on the critical endpoint location in continuum QCD. 

Supplementing the analysis with the known heavy-ion chemical freeze-out
line~(Fig.~\ref{fig:exclProb}), we can exclude the existence of the critical
point, at the 2$\sigma$ level, below $\mu_B^{2\sigma}=450$~MeV.  Our results
for the critical endpoint are consistent with the conclusions from an analysis
of the Polyakov loop in a small physical volume in
Ref.~\cite{Borsanyi:2024xrx}, where a broadening of the crossover transition
was observed at $\mu_B \lesssim 400 \MeV$.

From chiral effective models one expects the transition temperature in the
chiral limit at zero density to be an upper bound for the critical point at
finite density \cite{Halasz:1998qr}.  The O(4) critical temperature in the
chiral limit was recently calculated to be $T_{\rm \chi} \sim
133$~MeV~\cite{HotQCD:2019xnw,Kotov:2021rah}. Translating our lower bound on
the critical chemical potential to an upper bound in temperature, our finding
is consistent with the above expectation.

\bigskip

\emph{Acknowledgments} --
This work is supported by the MKW NRW under the funding code NW21-024-A.
Further funding was received from the DFG under the Project No. 496127839.
This work was also supported by the Hungarian National Research, Development
and Innovation Office, NKFIH Grant No. KKP126769.  This work was also supported
by the NKFIH excellence grant TKP2021{\textunderscore}NKTA{\textunderscore}64.
This work is also supported by the Hungarian National Research, Development and
Innovation Office under Project No. FK 147164.  This material is also based
upon work supported by the National Science Foundation under grants No.
PHY-2208724, and PHY-2116686, and within the framework of the MUSES
collaboration, under grant number No. OAC- 2103680. This material is also based
upon work supported by the U.S. Department of Energy, Office of Science, Office
of Nuclear Physics, under Award Numbers DE-SC0022023 and DE-SC0025025, as well
as by the National Aeronautics and Space Agency (NASA)  under Award Number
80NSSC24K0767.  The authors gratefully acknowledge the Gauss Centre for
Supercomputing e.V. (\url{www.gauss-centre.eu}) for funding this project by
providing computing time on the GCS Supercomputer HAWK at HLRS, Stuttgart.  An
award of computer time was provided by the INCITE program. This research used
resources of the Argonne Leadership Computing Facility, which is a DOE Office
of Science User Facility supported under Contract DE-AC02-06CH11357.

\bibliography{apssamp}

\pagebreak
\widetext
\begin{center}
\textbf{\large Supplemental material}
\end{center}
\setcounter{equation}{0}
\setcounter{figure}{0}
\setcounter{table}{0}
\setcounter{page}{1}
\makeatletter
\renewcommand{\theequation}{S\arabic{equation}}
\renewcommand{\thefigure}{S\arabic{figure}}
\renewcommand{\thetable}{S\arabic{table}}

\section{Equation of state at $\mu_B=0$}

\subsection{Trace anomaly}

\begin{figure}[b]
    \centering
    \includegraphics[width=0.75\linewidth]{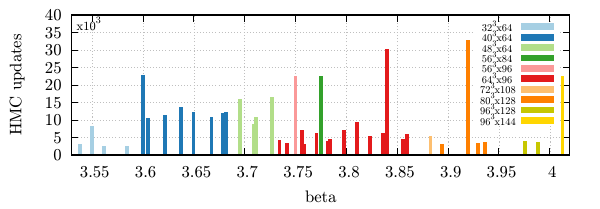}
    \caption{Accumulated statistics for the vacuum runs used for renormalization. Different colors indicate different lattices.}
    \label{fig:vacuumruns}
\end{figure}

We performed high-statistics simulations dedicated to the renormalization of the gauge action. We show in Fig.~\ref{fig:vacuumruns} the statistics we collected for the simulated values of $\beta$.

The trace anomaly is defined as:
\begin{equation}
\frac{I(T)}{T^4} = - T N_\tau^4 \frac{d \beta}{d T} \left( \left\langle - s_G \right\rangle + \sum_f \frac{dm_f}{d\beta} \left\langle \bar{\psi}_f \psi_f  \right\rangle \right) \, \, ,
\end{equation}
where $d\beta/dT$ is given by the scale setting and the $dm_f/d\beta$ 
are given by the line of constant physics.
The expectation values $\left\langle - s_G \right\rangle$ and 
$\left\langle \bar{\psi}_f \psi_f  \right\rangle$ are renormalized by subtracting 
the vacuum contribution at the same bare parameters:
\begin{align}\label{smeq:rensG}
\left\langle - s_G \right\rangle &= \left\langle - s_G \right\rangle_T - \left\langle - s_G \right\rangle_0 \, \, , \\\label{smeq:renpbp}
\left\langle \bar{\psi}_f \psi_f  \right\rangle &= \left\langle \bar{\psi}_f \psi_f  \right\rangle_T - \left\langle \bar{\psi}_f \psi_f  \right\rangle_0 \, \, .
\end{align}

In order to determine the continuum limit of $I(T)$, we employ a procedure whereby 
we separately fit the vacuum gauge action, and the fixed-$N_\tau$ trace anomaly 
\textit{minus} the aforementioned vacuum gauge action.

For the first fit, the chi square reads:
\begin{align}
\chi^2_0 &= \sum_{m \in \Lambda_0} \frac{\left( \sum_{a=0}^{d} \alpha_a \beta^a -  \left\langle -s_g \right\rangle_0 \right)^2}{\sigma_m^2} \, \, ,
\end{align}
where the vacuum gauge action, common to all lattices, is modeled with a polynomial 
of order $d$ in $\beta$:
\begin{equation}
\left\langle - s_g \right\rangle_0 = \sum_{a=0}^{d} \alpha_a \beta^a \, \, ,
\end{equation}
and with $\sigma_m^2$ the errors on $\left\langle -s_g \right\rangle_0$.

For the second fit, the chi square reads:
\begin{align}
\chi^2_T &= \sum_{n \in \Lambda_T} \frac{\left( \sum_{i \in \rm K} S_i (T) \left( c_i + \frac{d_i}{N_\tau^2}\right) + T N_\tau^4 \frac{d\beta}{dT} \sum_{a=0}^{d} \alpha_a \beta^a -  g_n \right)^2}{\sigma_n^2} \, \, ,
\end{align}
where the fixed-$N_\tau$ trace anomaly \textit{minus} the vacuum gauge action is:
\begin{equation}
g_n = T N_\tau^4 \left[ \left\langle - s_g \right\rangle_T \frac{d\beta}{dT} + \sum_{f \in \rm flavours} \left\langle \bar{\psi}_f \psi_f \right\rangle \frac{dm_f}{dT} \right] \, \, ,
\end{equation}
with $\sigma_n^2$ the errors on $g_n$. We model this quantity with cubic splines
$S_i(T)$ with nodes K, where each parameter is additionally a linear model in 
$1/N_\tau^2$.

\begin{figure}
    \centering
    \vspace{2mm}
    \includegraphics[height=0.34\linewidth]{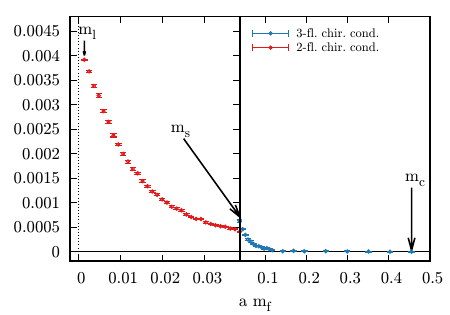}
    \hfill
    \includegraphics[height=0.34\linewidth]{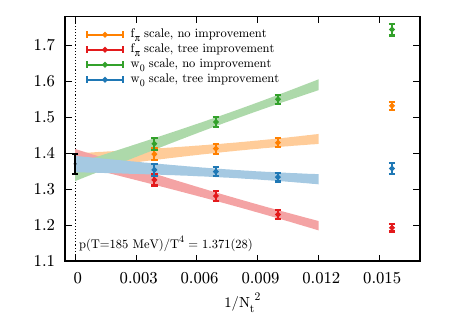}
    \caption{Left: chiral condensate in the $2-$ and $3-$flavour theories for quark masses between the light $m_l$ and charm $m_c$ values. Right: continuum limit of the pressure obtained directly at $T=185 \MeV$ from Eq.~\eqref{smeq:pconst}.}
    \label{fig:intconstant}
\end{figure}

We perform the global fit by minimizing the combined chi square:
\begin{align} \label{smeq:chisq_mega}
\chi^2 &= \chi^2_0 + \chi^2_T \, \, .
\end{align}

We consider several sources of systematic uncertainties:
\begin{itemize}
    \item we set the scale with either $f_\pi$ or $w_0$;
    \item we employ fits of order $d=5,6$ for the vacuum gauge action;
    \item we use two different fit ans\"atze for the $\bar\psi\psi$ renormalization;
    \item we employ two different sets of node points for the spline fits;
    \item we apply, or not, the tree level improvement to the finite-$N_\tau$ trace anomaly.
\end{itemize}
These choices give us 32 analyses which we combine with the histogram method of 
Ref.~\cite{Borsanyi:2020mff}. 

\subsection{Integration constant}

In this work we choose to calculate the integration constant for the pressure at 
$T_0 = 185 \MeV$, to minimize the uncertainty on the equation of state in 
the transition region. The pressure is obtained as an integral of the renormalized chiral 
condensates. We performed dedicated simulations at $T_0 = 185 \MeV$ on lattices with 
$N_\tau=8,10,12,16$ and aspect ratio $LT=4$. We start the integration from the mass 
of the charm quark $m_c$, integrate the 3-flavour theory down to the strange quark 
mass $m_s$, then integrate the two-flavour theory down to the light quark mass $m_l$: 
\begin{align} \label{smeq:pconst} 
\frac{p(T_0)}{T_0^4} &= \int_{m_s}^{m_l} dm_2 \pbp{\!}_{2} \,  (m_2) 
+ \int_{m_c}^{m_s} dm_3 \pbp{\!}_{3} \, (m_3) \, \, .
\end{align}
In the left panel of Fig.~\ref{fig:intconstant} we show the integrands as
functions of the quark mass in the 2- (red, left) and 3- (blue, right) flavour
theories, on our $48^3\times 12$ lattice. The resulting total integral will be
$2\times$ the integral of the red curve plus $3\times$ the integral of the blue
one. We note that these data, too had to be generated for all four of our
lattice spacings, and for every mass $T=0$ simulations it was necessary to
calculate the additive counterterm to $\bar\psi\psi$.  While for $m<m_s$ these
vacuum contributions could be easily interpolated (we employed a 2/2 rational
function), for $m\ge m_s$ each value of $m$ and $N_\tau$ needed a dedicated
renormalization run. For $m>m_s$ we fitted the exponential of a 1/1 rational
function and integrated that.

We also varied the scale setting for the integration constant. If the
temperature is fixed at $T_0=185~\mathrm{MeV}$, a different scale setting would
require a different gauge coupling $\beta$, with new renormalization runs, etc.
Instead of simulating scale setting specific, and expensive ensembles, we
calculated the scale setting corrections by integrating the trace anomaly
between the temperatures that belong to the simulated bare parameters with the
respective scale setting. This temperature interval is $N_\tau$ dependent,
thus, the continuum extrapolations have a very different slope as we switch
between scale setting parameters (see right panel of
Fig.~\ref{fig:intconstant}).  In the same plot we also include the
extrapolation with and without tree level improvement to the finite-$N_\tau$
pressure. 

The final results for the equation of state at zero chemical potential are listed in Table~\ref{tab:EoS}.

\input{tab.tex}

\section{Entropy contours for $\mu_B \neq 0$}
The calculation of the entropy contours is done in two steps. In the first
step, the entropy is determined as a function of $T$ and purely imaginary
$\mu_B$ in full continuum QCD. The second step is the analytic continuation of
these contours to real chemical potentials.

\subsection{Calculation of entropy at purely imaginary chemical potentials}
The determination of the entropy as a function of temperature and imaginary chemical potential 
starts by measuring the first derivative 
\begin{equation}
\hat n_1 = \left( \frac{\partial (p/T^4)}{\partial (\mu_B/T)} \right)_{n_S=0}\rm.
\end{equation}
Note that the chemical potential derivative is taken along the direction defined by vanishing strangeness density and electric charge chemical potential $\hat n_1(T,\mu_B/T) = n_B(T,\mu_B)/T^3$. 

\begin{figure}
    \centering
    \includegraphics[width=0.48\linewidth]{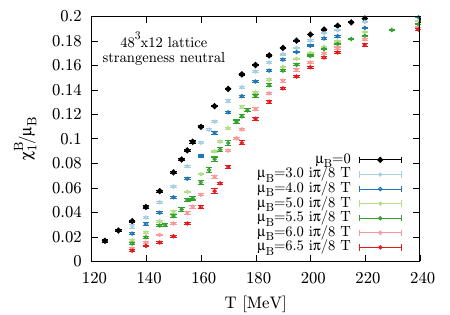}
    \includegraphics[width=0.48\linewidth]{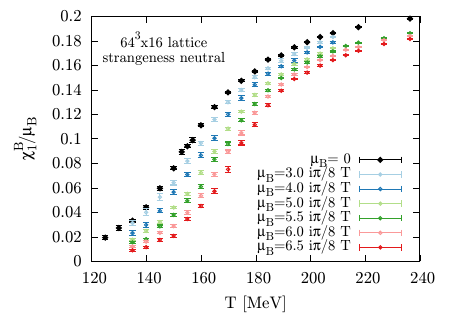}
    \caption{\label{fig:immudata}
    Lattice data of $\hat n_1(T)/\hat \mu_B$ as a function of the temperature for the simulated
    imaginary (and zero) values of chemical potentials. We show the data on our two finest lattices:
    $48^3\times 12$ and $64^3\times 16$.
    }
\end{figure}

In Fig.~\ref{fig:immudata} we show the raw input data as obtained from the
simulations, for our two finest lattices. We generated 4000 configurations per
ensemble.  For this work we generated a data set at $\mu_B/T=5.5i \pi/8$ on the
$40^3\times10$ lattice, and at $\mu_B/T=\left(5.5, 6 \,\mathrm{and}
\,6.5\right) i\pi/8$ on the $64^3\times16$ lattice.  On the $48^3\times 12$
lattice, all these were already available and used e.g. in
Refs.~\cite{Borsanyi:2020fev,Borsanyi:2022qlh}.

Within each ensemble we computed $\hat n_1$, the strangeness density as well
the $\mu_S$-derivative of both. The parameters were previously tuned such that
the strange expectation value approximately vanishes.  We used the $\mu_S$
derivatives to extrapolate $\hat n_1$ to exactly vanishing strangeness. Tuning
to strangeness neutrality was discussed in more detail in
Refs.~\cite{Borsanyi:2020fev, Borsanyi:2022qlh}.

After the calculation of $\hat n_1$ we utilize the following thermodynamics relations:
\begin{equation}
s(T, \mu_B) = \left(\frac{\partial p}{\partial T}\right)_{\mu_S} \,  , \quad\quad
p(T, \mu_B) = \int_0^{\mu_B} n_B(T,\mu_B^{\prime}) d\mu_B^{\prime} \rm,
\end{equation}
where $p$ is the pressure. 
Converting derivatives at fixed $\mu_B$ to derivatives at fixed $\hat \mu_B=\mu_B/T$ we get~\cite{Borsanyi:2021sxv}:
\begin{equation}
\label{smeq:entropy_mupos}
\begin{aligned}
\hat s(T, \hat \mu_B) - 
\hat s(T, 0) =  
4\left( \hat p(T, \hat \mu_B) - 
\hat p(T, 0) \right) +  
 \left( \frac{\partial \hat p(T, \hat \mu_B)}{\partial T} - 
\frac{\partial \hat p(T, 0)}{\partial T} \right) - \hat \mu_B \hat n_1
  \rm,
\end{aligned}
\end{equation}
where the dimensionless pressure is $\hat p = p/T^4$ and the dimensionless
entropy is $\hat s = s/T^3$.  In order to perform the $T$ derivatives and
$\mu_B$ integrals needed for the calculation of the entropy, as well as for the
continuum limit extrapolation, we fit $\hat n_1$ with the following linear
ansatz:
\begin{equation}
\hat n_1 (T, \hat \mu_B) = \sum_{klm} c_{klm} f_k(T) g_l(\hat \mu_B) h_m(N_t) \rm,
\end{equation}
where the $c_{klm}$ are fitted real parameters, while the $f_k(T)$, $g_l(\hat
\mu_B)$ and $h_m(N_t)$ are fixed basis functions in the $T$, $\hat \mu_B$ and
$N_t$ directions respectively. By exactly differentiating the basis function
$f_k(T)$ and exactly integrating the basis functions $g_l(\hat \mu_B)$ the
different terms in equation~\eqref{smeq:entropy_mupos} can be obtained. For the
different basis functions in the $N_t$ direction we choose $h_0=1$ , $h_1 =
1/N_t^2$. Thus, we perform a linear continuum extrapolation in $1/N_t^2$. Note
that we always include a tree level correction in the observable $\hat n_1$.
In the chemical potential direction we have {2 different choices} of the basis functions:
\begin{enumerate}
\item $g_0 = 1$, $g_1 = \hat \mu^2$, $g_2 = \hat \mu^4$, $g_3 = \hat \mu^6$, and
\item $g_0 = 1$, $g_1 = \hat \mu^2$, $g_2 = \hat \mu^4$, $g_3 = \hat \mu^6$, $g_4 = \hat \mu^8$\rm.
\end{enumerate}
In the $T$ direction we use cubic splines, with fixed node points at several temperatures. To estimate the systematics of this temperature interpolation we use {4 different choices} for the node points:
\begin{enumerate}
\item $[129, 142, 152, 162, 172, 182, 192, 202, 222,251]$~MeV,
\item $[129,145, 155, 165, 175, 185, 195, 205, 225,251]$~MeV,
\item $[129, 139, 148, 158, 168, 178, 188, 198, 218,238, 251]$~MeV and
\item $[129,144, 159, 174, 189, 204, 219, 234, 251]$~MeV.
\end{enumerate}
Another source of systematic errors that is taken into account at this point are {2 different choices} for the scale setting: either the pion decay constant $f_\pi$ or the 
Wilson flow scale $w_0$ introduced in Ref.~\cite{BMW:2012hcm}. 

Finally, the finite chemical potential contribution of Eq. \eqref{smeq:entropy_mupos} is added to the $\mu_B=0$ contribution, and the results are then utilized for the final step of this analysis, the analytic continuation step.

\subsection{Entropy contours}

The entropy contours are modeled as 1/1 rational functions in $\mu_B^2$. 
Data points at different imaginary $\mu_B$ are correlated, since the entropy itself is a
result of an integration in imaginary $\mu_B$. If the $p$-value of the correlated fit is
below a cut (0.02), we drop that $T_0$ value from the analysis. 

For a fixed, positive $\mu_B^2$ parameter we consider several contours,
defined by the axis intercept $T_0$ in 1 MeV steps in the range $131\dots 180$~MeV.
The $T_s$ contour value as a function of $T_0$ is the basis of the further analysis.

This $T_s(\mu_B^2,T_0)$ function is available to us numerically with jackknife errors.
Whether the entropy is multi-valued at a given point, is determined on the basis of the $\partial T_s/\partial T_0$
derivative. This requires some level of interpolation. We use fourth degree polynomials
to find the derivatives. We perform fully correlated fits to the data points.

As systematics we vary the following options:
\begin{itemize}
    \item the upper range of the fit $T_{0,\rm max}= 161$~or $170$~MeV;
    \item the lower range of the fit $T_{0,\rm max}= 130$~or $135$~MeV;
    \item we sample $T_0$ every 2,3 or 4~MeV
\end{itemize}
For the full analysis, more than 16000 such fits are performed if we count all $\mu_B$ values, where we
step in 10 MeV units. 49.3 \% of the fits have a $p$-value better than 0.5.

We work through the temperatures in the phase diagram in 2 MeV steps.
For each selected $T$ we solve $T_s(\mu_B^2,T_0)=T$, thus arriving at a $T_0$ value with some
statistical error. This error is propagated to $\partial T_s/\partial T_0$.

Since all considered fits were acceptable, we use uniform weighting when we
combine the analyses.  In practice, we average Gaussian cumulative distribution
functions defined by the central sample and the jackknife error of $\partial
T_s/\partial T_0$. The resulting distribution allows us to identify those
positions in the phase diagram where the derivative is inconsistent with zero
or negative values with a confidence level that would correspond to $2\sigma$
for a Gaussian distribution.

\end{document}

%% file: tab.tex
\begin{table}
\begin{tabular}{|c|r|r|r|r|}
\hline
T~[MeV] & $p(T)/T^4$ & $s(T)/T^3$ & $\epsilon(T)/T^4$ & $I(T)/T^4$  \\
\hline
120 & \quad 0.433(53) \quad & \quad 2.41(25) \quad & \quad 1.98(22) \quad & \quad 0.69(18) \quad \\ 
125 & \quad 0.464(51) \quad & \quad 2.68(18) \quad & \quad 2.22(15) \quad & \quad 0.83(13) \quad \\ 
130 & \quad 0.500(47) \quad & \quad 2.99(18) \quad & \quad 2.49(16) \quad & \quad 0.99(19) \quad \\ 
135 & \quad 0.541(43) \quad & \quad 3.34(21) \quad & \quad 2.80(21) \quad & \quad 1.18(23) \quad \\ 
140 & \quad 0.587(40) \quad & \quad 3.76(23) \quad & \quad 3.18(22) \quad & \quad 1.42(22) \quad \\ 
145 & \quad 0.642(39) \quad & \quad 4.30(21) \quad & \quad 3.66(20) \quad & \quad 1.74(19) \quad \\ 
150 & \quad 0.707(38) \quad & \quad 4.96(16) \quad & \quad 4.25(14) \quad & \quad 2.14(13) \quad \\ 
155 & \quad 0.784(36) \quad & \quad 5.69(13) \quad & \quad 4.90(11) \quad & \quad 2.55(11) \quad \\ 
160 & \quad 0.871(34) \quad & \quad 6.39(14) \quad & \quad 5.52(12) \quad & \quad 2.91(12) \quad \\ 
165 & \quad 0.965(32) \quad & \quad 7.05(14) \quad & \quad 6.08(12) \quad & \quad 3.19(13) \quad \\ 
170 & \quad 1.063(30) \quad & \quad 7.66(14) \quad & \quad 6.59(13) \quad & \quad 3.40(13) \quad \\ 
175 & \quad 1.164(29) \quad & \quad 8.23(15) \quad & \quad 7.06(14) \quad & \quad 3.57(14) \quad \\ 
180 & \quad 1.267(28) \quad & \quad 8.78(17) \quad & \quad 7.51(15) \quad & \quad 3.71(14) \quad \\ 
185 & \quad 1.371(28) \quad & \quad 9.32(18) \quad & \quad 7.95(16) \quad & \quad 3.83(14) \quad \\ 
190 & \quad 1.474(28) \quad & \quad 9.84(19) \quad & \quad 8.36(17) \quad & \quad 3.93(14) \quad \\ 
195 & \quad 1.578(29) \quad & \quad 10.32(21) \quad & \quad 8.74(19) \quad & \quad 4.00(16) \quad \\ 
200 & \quad 1.680(30) \quad & \quad 10.77(23) \quad & \quad 9.09(21) \quad & \quad 4.04(16) \quad \\ 
205 & \quad 1.780(32) \quad & \quad 11.17(24) \quad & \quad 9.39(22) \quad & \quad 4.05(16) \quad \\ 
210 & \quad 1.878(33) \quad & \quad 11.53(24) \quad & \quad 9.65(22) \quad & \quad 4.02(15) \quad \\ 
215 & \quad 1.972(35) \quad & \quad 11.86(24) \quad & \quad 9.89(21) \quad & \quad 3.97(14) \quad \\ 
220 & \quad 2.062(36) \quad & \quad 12.16(25) \quad & \quad 10.10(22) \quad & \quad 3.91(14) \quad \\ 
225 & \quad 2.150(38) \quad & \quad 12.45(25) \quad & \quad 10.30(22) \quad & \quad 3.85(14) \quad \\ 
230 & \quad 2.234(39) \quad & \quad 12.72(26) \quad & \quad 10.49(23) \quad & \quad 3.79(14) \quad \\ 
235 & \quad 2.315(41) \quad & \quad 12.99(27) \quad & \quad 10.67(23) \quad & \quad 3.73(14) \quad \\ 
240 & \quad 2.392(42) \quad & \quad 13.24(27) \quad & \quad 10.85(23) \quad & \quad 3.67(14) \quad \\ 
245 & \quad 2.467(44) \quad & \quad 13.48(28) \quad & \quad 11.01(24) \quad & \quad 3.61(13) \quad \\ 
250 & \quad 2.540(46) \quad & \quad 13.71(29) \quad & \quad 11.17(24) \quad & \quad 3.55(13) \quad \\ 
255 & \quad 2.609(47) \quad & \quad 13.93(29) \quad & \quad 11.32(25) \quad & \quad 3.50(13) \quad \\ 
260 & \quad 2.675(49) \quad & \quad 14.14(30) \quad & \quad 11.46(25) \quad & \quad 3.44(13) \quad \\ 
265 & \quad 2.740(51) \quad & \quad 14.34(31) \quad & \quad 11.60(26) \quad & \quad 3.38(13) \quad \\ 
270 & \quad 2.803(53) \quad & \quad 14.54(31) \quad & \quad 11.74(26) \quad & \quad 3.33(13) \quad \\ 
275 & \quad 2.863(54) \quad & \quad 14.73(32) \quad & \quad 11.86(27) \quad & \quad 3.27(12) \quad \\ 
280 & \quad 2.922(56) \quad & \quad 14.91(33) \quad & \quad 11.99(27) \quad & \quad 3.22(13) \quad \\ 
285 & \quad 2.978(58) \quad & \quad 15.08(34) \quad & \quad 12.10(28) \quad & \quad 3.17(14) \quad \\ 
290 & \quad 3.033(59) \quad & \quad 15.25(35) \quad & \quad 12.21(30) \quad & \quad 3.11(15) \quad \\ 
295 & \quad 3.086(61) \quad & \quad 15.40(37) \quad & \quad 12.32(32) \quad & \quad 3.06(18) \quad \\ 
\hline 
\end{tabular} 
\caption{ 
\label{tab:EoS}
The tabulated equation of state at zero chemical potential: pressure, entropy density, energy density and trace anomaly.
}
\end{table}